\documentclass[12pt]{article}

\usepackage{times}

\usepackage{graphicx}

\begin{document}

\title{\textbf{Social Effects in Science}\\
Modelling Agents for a Better Scientific Practice}

\author{Andr\'e C. R. Martins\\
GRIFE -- EACH\\
Universidade de S\~ao Paulo\\
S\~ao Paulo, Brazil\\
}
\date{}
\maketitle

\begin{abstract}
  Science is a fundamental human activity and we trust its results because it has several error-correcting mechanisms. Its is subject to experimental tests that are replicated by independent parts. Given the huge amount of information available, scientists have to rely on the reports of others. This makes it possible for social effects to influence the scientific community. Here, an Opinion Dynamics agent model is proposed to describe this situation. The influence of Nature through experiments is described as an external field that acts on the experimental agents. We will see that the retirement of old scientists can be fundamental in the acceptance of a new theory. We will also investigate the interplay between social influence and observations. This will allow us to gain insight in the problem of when social effects can have negligible effects in the conclusions of a scientific community and when we should worry about them. 
\end{abstract}

\section{Introduction}

Science is a human endeavor built with a powerful system of error-correcting techniques. Ideally, the quality of its conclusions are judged not by any social measures, but by how well its predictions match the observations of the real world. Experiments are checked and replicated and, even though errors can and do happen, they are corrected with time. Such is the picture that we like to believe in. And, for several areas of knowledge, this seems to be a very accurate description of scientific activity.

However, there are people who propose a very different picture of how ideas are accepted in Science. The Strong Program in the Sociology of Scientific Knowledge, among other accounts \cite{barnesbloor82a,pickering84a,latourwoolgar86a,bloor91a}, defends that all scientific knowledge is originated only from social effects. Taken to the ultimate consequences, that would mean our current knowledge is no better than old myths, legends or pseudo-science.

While these descriptions do go too far on the consequences of social influence in the conclusions of a scientific community, it is true that real scientists are subject to social pressures. Even worse, social effects are often unavoidable, as the only way we can work and arrive at strong conclusions is by allowing other people to convince us, at least in fields where we are not experts. Arguments of authority are unavoidable \cite{hardwig85a}, since the current body of knowledge is too big for a single human mind. This can be true even in descriptions of one single experiment. ``Big Science'' projects, as, per example the Human Genome project, involve hundreds of scientists and technicians who must trust the information provided by their fellows, instead of checking it all by themselves.

Therefore, we need to evaluate the impact of social influence  on the convergence to the best description \cite{goldman87a}. Accounts that incorporate the social aspects of the practice of Science and show us how we can still expect that the best description be chosen do exist \cite{kitcher93a}, but efforts in this direction are still very rare. We need to develop tools that will help us to explore the circumstances when social influence can cause a community to choose worse theories. That way, our confidence in Science, in the cases where it is warranted, will only become stronger and better justified. 

Since we are talking about a scientific community, it is natural to look for tools that can describe the beliefs of a community as a consequence of the beliefs of individual agents. So far, simulation of scientific communities is a field of study that has been mostly applied to describing the dynamics of things like papers and ideas \cite{newman01a,borneretal04a,newman08a}, but that is not the approach here. Instead, we can use the tools of Opinion Dynamics \cite{castellanoetal07,galametal82,deffuantetal02a,hegselmannkrause02,galam05b}, that describe the general social properties of the spread of opinions, as a basis for describing an artificial society of scientific agents.

In this paper, a model for an artificial scientific community will be presented. The consequences of the model for the appearance of a consensus in favor of the the true description will be studied and we will investigate the conditions under which a paradigm shift can be observed in the artificial society. Both the consequences of the particular model presented here and the general idea of using agent models as a basis for a better epistemic practice will be discussed.

\section{Artificial Scientific Agents}

A real scientist, when analyzing which theory better describes her field of expertise, will probably use a large amount of data and information, carefully weighted. If the subject she is evaluating is just related to the area she works with, her evaluation will almost certainly be simpler, involving less data. Real evaluations can be quite complicated. However, our aim at this work is not describing real scientists. Like the Ising model for spins in Physics, the idea is to use a simplified version of the behavior of scientists in order to explore the consequences in the macro scale of the society. That is, we aim at generality of conclusions, at the expense of precise predictions. Yet, some resemblance with a correct reasoning is, of course, a desirable property of the model.

This can be achieved by using Bayesian rules as a basis for generating Opinion Dynamics problems \cite{martins08e}. The problem we are interested in, of choosing a theory between two competing alternatives, $A$ and $B$, can be represented as a discrete choice between those alternatives. In order to make the model more realistic, instead of spins, the internal opinion of each scientist can be associated with a probability $p_i$ that each artificial scientist $i$ assigns to the possibility that $A$ is the best description. This description corresponds to the ``Continuous Opinions and Discrete Actions'' model (CODA model), previously used to explore the emergence of extremism in artificial societies \cite{martins08a,martins08b}. This model has been previously applied with success to describe the time of adoption of innovations \cite{martinspereira08a}, a problem similar to the adoption of new ideas and should provide a good basis for modeling a scientific community.

The model is based on the idea that, if $A$ is right, each of the neighbors of an agent, located in a given social network \cite{newmanetal06a,vegaredondo2007a}, will have a probability $\alpha>0.5$ of choosing $A$(and similarly, for $B$). In this context, it is easier to work the quantity
\[
\nu_i=\ln\frac{p_i}{1-p_i}.
\]
Here, if $\nu_i>0$, we have $p_i>0.5$ and, therefore, a subjective belief in favor of $A$; if $\nu_i<0$, the agent chooses $B$. By applying Bayes Theorem, we obtain a very simple update rule for $\nu_i$, when agent $i$ observes the choice of its neighbor $j$, given by
\[
\nu_i (t+1)= \nu_i (t)+ \textnormal{sign}(\nu_j)*a,
\]
where $a$ is a step size that depends on how likely the agents believe it is that their neighbors will be correct, that is, it is a function of $\alpha$. If we renormalize the update rule, by using $\nu_i^*=\nu_i /a$ instead, we will have
\begin{equation}
\nu_i^* (t+1)= \nu_i^* (t)+ \textnormal{sign}(\nu_j^*),
\end{equation}
making it clear that the value of $a$ is irrelevant to the dynamics of choices, since that dynamics depends only on the signs of $\nu_i$ (or $\nu_i^*$). 

Unlike a spin model, as $\nu_i^*$ moves away from zero, it becomes more difficult for an agent to change its choice, since it now has a stronger opinion. When agents follow this simple dynamics on a quenched network, local communities reinforce their choices and we observe the survival of both choices in the long run. The opinions become more extreme at each interaction and extremists are observed in the interior of each domain. Notice that the influence comes only as an observation of the neighbor preference. No reports or arguments are described, in order to obtains a simplified version of the social influence process.

Under these rules, the opinion of the agents is updated only due to social effects. In order to model a scientific community, we need to describe the influence of the real world, by means of experimental results. This can be easily accomplished by the introduction of an external field that points in the direction of the best theory. Here, we will assume, without loss of generality, that $A$ describes the world better. Of course, the agents do not know that.

First, we need to acknowledge that not every scientist makes experiments. Even those who do work in experimental problems might learn the results only from the report of others in their team. Therefore, in the model, only a fraction $\tau$ of the agents actually interact with the external field (that always favors $A$). Assuming that an agent is an experimental agent and will be influenced by the field, we need to compare the importance of social effects and observations of the world. After all, even experimental scientists obtain information they use for their opinions from their peers. 

This effect can be introduced by assigning a relative strength that measure the importance of direct observations versus the opinions of other scientists. This was implemented in the model as a probability $\rho$ that, at each interaction, the agent would be influenced by the world instead of a neighbor agent. Interacting with a neighbor or the external field causes the same change in the opinion, as stated in Equation 1. This is not a problem since a larger $\rho$ can represent either a more frequent experimenting or less frequent experiments, but with more importance associated to the experiments than to the observation of opinions.

\begin{figure}[ht]
 \includegraphics[width=0.8\textwidth]{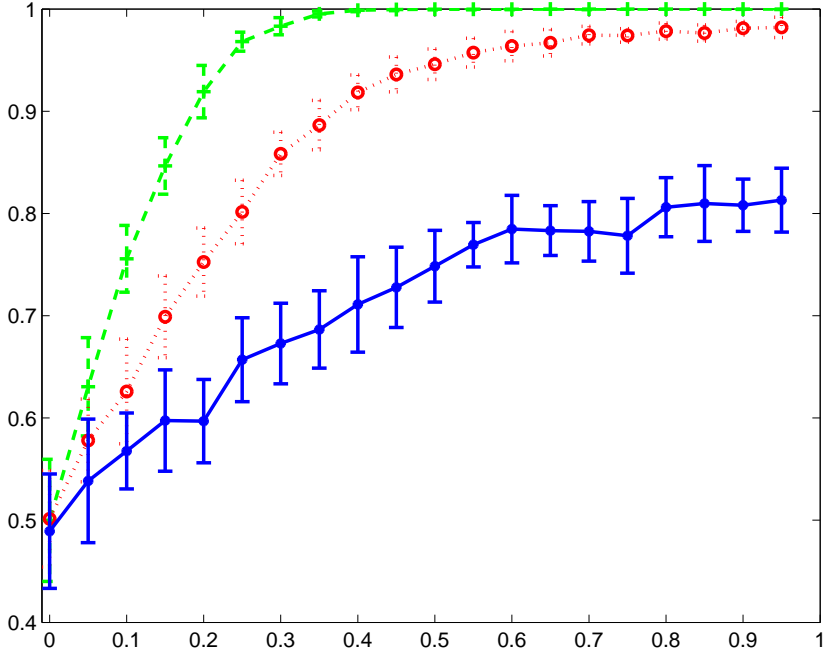}
 \caption{Proportion of agents aligned with the external field (correct scientists) as a function of $\rho$.  Every realization had initial conditions where each agent had 50\% chance of supporting either theory. The blue line corresponds to $\tau=0.1$, the red one to $\tau=0.25$,  and the green one to $\tau=0.5$.
}\label{fig:externalvariable}
 \end{figure}

Figure 1 shows the final proportion of agents after the choices freeze due to the strengthening of the opinions. A square bi-dimensional regular lattice with periodic boundary conditions and $N=32^2$ agents was used, where each agent only interacted with its four first neighbors (von Neumann neighborhood). As initial condition, every agent had an initial opinion $\nu_i^*$ less than one, in module, so that one interaction could change its choice. Also, the sign of $\nu_i$ was chosen with 50\% chance associated to each choice. The blue line corresponds to $\tau=0.1$, the red one to $\tau=0.25$,  and the green one to $\tau=0.5$. The central points are averages over 20 different realizations and the error bars are the observed standard deviation of the observed proportions

We can see that, except for $\tau=0.5$, a proportion of agents who do not choose $A$ can always be observed regardless of the value of $\rho$. For $\rho>0.5$, the experimental agents will always be convinced of the truth of $A$, since they give more importance to their own observations than social effects. But we still observe clusters of non-experimental agents who are able to keep their wrong views due to social reinforcement.

\begin{figure}[ht]
 \includegraphics[width=0.8\textwidth]{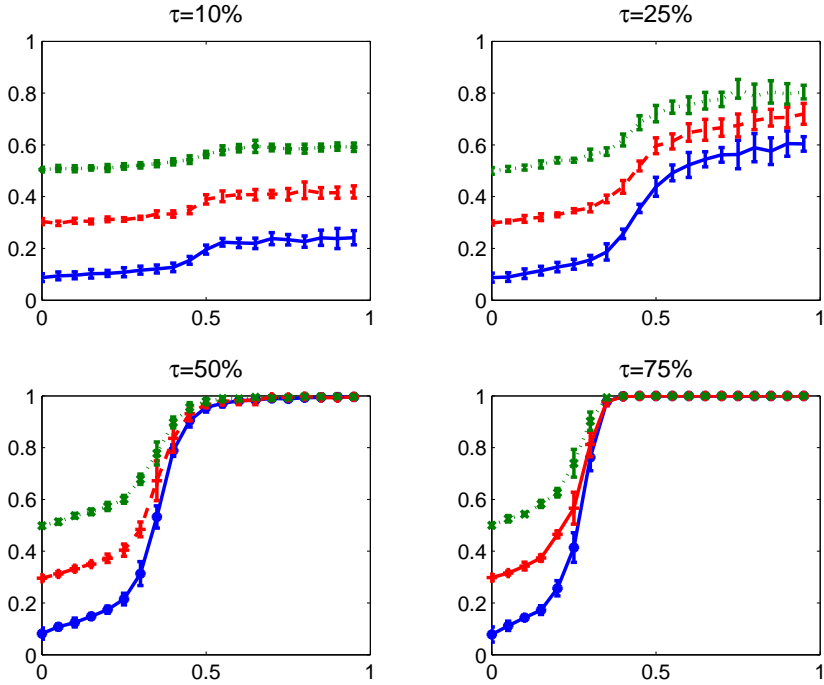}
 \caption{Proportion of agents aligned with the external field (correct scientists) as a function of $\rho$, for different proportions $\tau$ of agents who are influenced by the external field. The different curves correspond to different initial conditions. The blue line shows the case where 10\% of the agents were initially aligned with the external field, the red line, 25\%, and the green one, 50\%.
}\label{fig:n32pexp10adopt}
 \end{figure}

Of course, when a new theory is created, most of the agents will not favor it at first. Figure 2 shows the results of simulations for different initial proportions of agents who support that correct theory $A$. It is clear that, for the smaller values of $\tau$, that is, for smaller proportion of experimental agents, consensus is, once again, never reached. The proportion of agents who choose $A$ grows with $\rho$ but, unless there are enough agents who perform their own experiments, consensus is not achieved, regardless of the importance that experimenters assign to their own observations.

This is troublesome, since we would like to believe that the community should be able to change their minds due to the influence of the real world. However, social effects, here, are capable of preventing that convergence from happening. 

\section{Birth and Death}

What we have observed is that, as time passes, the local domains can survive and their opinions become even stronger. This tendency to extremism is a common feature of the CODA model without external fields and it is the reason that consensus is prevented. Its is interesting to notice that this correspond to the description that, quite often, old scientists do not change their minds when a new revolutionary theory appears (a paradigm shift) \cite{kuhn}. Instead, what allows the change of opinion in the community is the retirement and the death of the older, stubborn scientists.

\begin{figure}[ht]
 \includegraphics[width=0.8\textwidth]{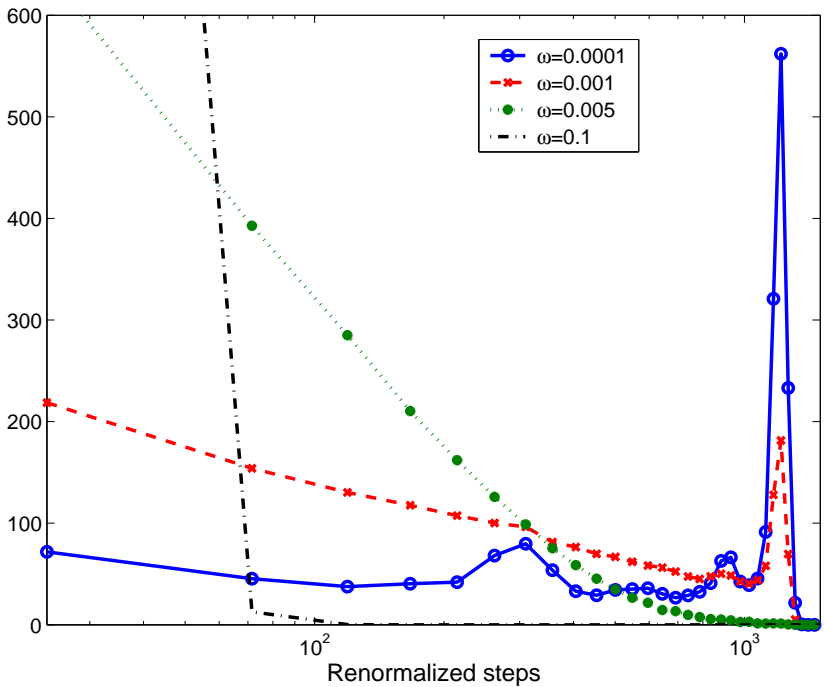}
 \caption{Distribution of renormalized opinions ($\nu_i^*$), for different probabilities of death, $\omega$.
}\label{fig:distn64birthdeathlog}
 \end{figure}
 
 It makes sense, therefore, to investigate the consequences of birth and death in the model. This will be done in a simple way. At each interaction, there will be a $\omega$ probability that a random agent will die. That is, $1/\omega$ measures the average number of interactions before an agent is replaced. When that happens, a new agent will appear at the same network location, but with moderate opinions equivalent to the initial conditions. Figure 3 shows, for the sake of comparison, the effect of introducing death in the original CODA model, when no external field is present (equivalent to $\tau=0$). 
 
 This is an interesting case per se, as it illustrates the effect of new generations in the problem of emergence of extremism. Notice that, for a low death rate, extremism is still the most important peak. As $\omega$ grows, an extremist tail is still observed, but the central moderate opinions become majority. One should notice, however, before arriving at unwarranted conclusions, that these are natural deaths, that don't influence the opinion of the neighbors of the dead agent. 
 
These results were obtained by assigning equal chances for the new agents to support each idea. This mechanism actually prevents the system from ever reaching consensus. Simulations with the probability of supporting $A$ for an agent born at time $t$ given by the proportion of agents that support $A$ at that instant showed that, for high death rates, the birth and death mechanism can, indeed, take the system to consensus, even in the absence of the external field.

\begin{figure}[ht]
 \includegraphics[width=0.9\textwidth]{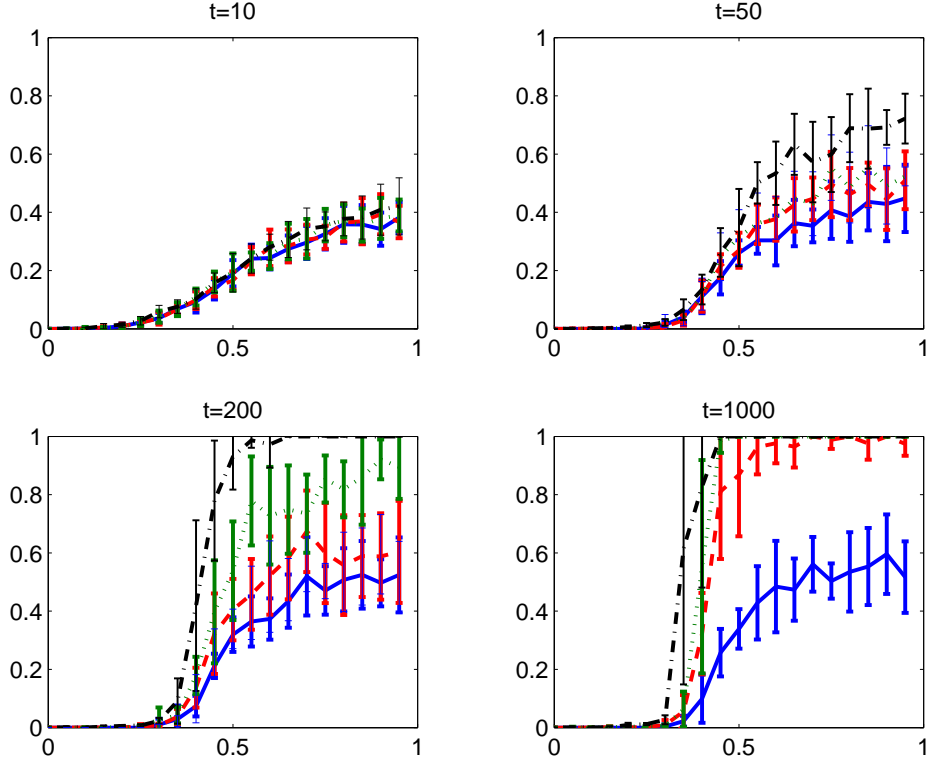}
 \caption{Proportion of agents aligned with the external field (correct scientists) as a function of $\rho$, for different values of $\omega$, after different number $t$ of average interactions per agent. No agent is influenced by the external field here ($\tau=0$). Blue lines show the results for $\omega=0$, the red ones, for $\omega=1\%$, the green ones, for $\omega=2\%$, and the black ones, for $\omega=5\%$,.
}\label{fig:difftdiffomegan16.eps}
 \end{figure}
 
It makes sense also that, when two theories are competing, a scientist that just enters the field will have an opinion that is more likely to favor the opinion of the majority. Therefore, from now on, new agents will support $A$ with a probability given by the proportion of current agents that do support $A$. The effects of introducing birth and death effects in the problem where there are experimental agents can be seen in Figures 4 and 5. In order to investigate the possibility of paradigm shifts, in every case, only 1\% of the agents supported theory $A$ at first. 

We can see that, as long as the experimental agents trust their own results enough, the replacement of scientists for new ones, with less extreme opinions, is beneficial towards the acceptance of the new theory by the community. This reinforces the idea proposed by Kuhn \cite{kuhn} that the retirement or death of old scientists is very important for the acceptance of new theories. 

However, we can also observe a few unexpected features of the model. If $\rho$ is small, no case showed the invasion of the new, better theory. Social effects were strong enough to prevent the theory from spreading and, in several realizations, it even caused the minority of supporters of $A$ to abandon their correct views.  Also, in the long run ($t=1,000$), the system presents a behavior that is typical of a first order phase transition, jumping from almost consensus against $A$ to consensus in favor of it.

\begin{figure}[ht]
 \includegraphics[width=0.9\textwidth]{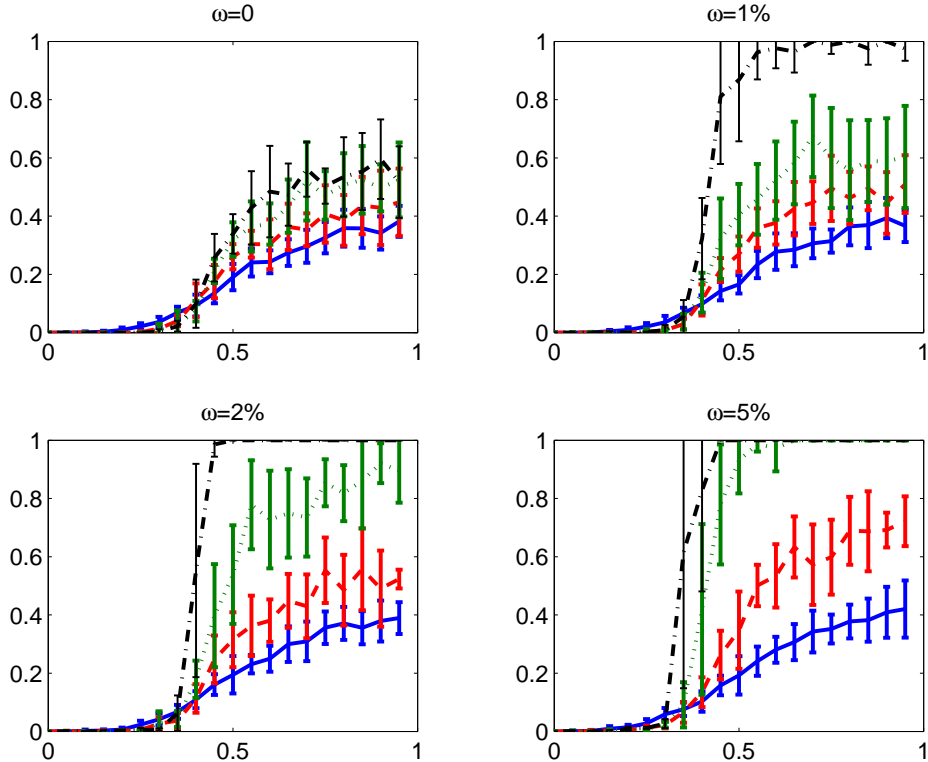}
 \caption{Proportion of agents aligned with the external field (correct scientists) as a function of $\rho$, for different values of $t$, with different values of $\omega$ in each graphic. Blue lines show the results for $t=10$ average interactions per agent, the red ones, for $t=50$,  the green ones, for $t=200$, and the black ones, for  $t=1.000$.
}\label{fig:diffomegadifftn16}
 \end{figure}
 
For $\rho >0.5$, it is clear that every experimental agent will be convinced about the superiority of $A$. However, the transition happens at lower values of $\rho$. Bellow $0.5$, the next important value of $\rho$ happens when, with one neighbor (possibly another experimental) supporting $A$, the experimental agent will tend to accept $A$ , despite the three contrary opinions. This will happen when
\[
\rho+(1-\rho)\frac{1-3}{4}=\frac{3\rho -1}{2}=0.
\]
that is, $\rho=1/3$. This value corresponds very well to the observed behavior. Longer simulations using larger lattices around $\rho=0.33$ showed that the system does seem to keep the consensus against $A$ for values of $\rho$ up to 0.33. As soon as we are above $1/3$, the proportion increases, with  large error bars showing that different realizations provide different proportions. As $\rho$ gets bigger, the system finally stabilizes at a consensus favoring the best theory. Although a further investigation of the limit with infinite agents could be interesting in order to decide if a real phase transition happens, a very large number of agents does not correspond to a real case. Communities of scientists investigating one specific theory are far from infinite.

The results presented here suggest that, as long as the social influence is not large enough (strong social influence correspond to small $\rho$), the system will eventually accept the correct theory. Different values of $\rho$ can lead to a faster or slower convergence but, as long as experimental agents give enough credit to their own observations, the agents will tend to the correct choice in time. 

Of course, those results assume that no error is ever made in the experiments and that the experiments always provide the right answer, regardless of statistical errors. These assumptions, depending on the circumstances, are strong ones. But relaxing these assumptions is beyond the scope of this paper and will be dealt with in future work. The main objective of the model was to show that Opinion Dynamics can make important contributions to epistemic problems. In an artificial world, we can decide which theory is the correct one and test strategies that allow the community to reach the correct consensus. With this, identifying potential problematic cases should become easier. Per example, for one single reader agent who makes no experiment, previous results show that replication plays a fundamental role on avoiding problems with non-identifiability of parameters, when error is supposed to exist \cite{martins08d}.

Of course, the model presented here is a simplification of actual scientific work and should not be used for quantitative predictions. However, it is interesting to notice that its qualitative description of the problem seems to match the existing accounts of scientific work. The importance of experiments was seen to be fundamental and the retirement of old scientists was observed to play a very important role in paradigm shifts. This illustrates the fact that this strategy of modeling scientific communities can really help us make better epistemic decisions and, eventually, find out which behaviors are more harmful to the community. Modeling scientists can help us make the results of Science more reliable.

A few speculations based on the results we have obtained are in order. Experiments are easier to conduct in the Natural Sciences. In Humanities, on the other hand, arguments of authority seem to be much more prevalent. That is, texts in Social Sciences, per example, often tell us how other people have made similar remarks about the problem before. This seems to indicate that social effects are much stronger in the Humanities than in the Natural Sciences. If that is true, the model presented here says that paradigm shifts towards better theories are to be expected in Physics and Biology, but not in Social Sciences. That is, scientific results would be based on reality in the first case, and social constructs in the second one. Therefore, it should be no surprise that analysis of Science by sociologists would conclude in favor of the social construction of ideas. Assuming the model is correct, that would be, indeed, the current state of their field. The good news is that this suggests a way to improve how Social Sciences are studied. Less importance to authority arguments seems like an important first change in order to give less importance to social effects and, therefore, allow reality to influence scientists better.

\section*{Acknowledgements} 
The author would like to thank Funda\c{c}\~ao de Amparo \`a Ci\^encia do Estado de S\~ao Paulo (FAPESP), under grant 2008/00383-9, and the Conselho Nacional de Pesquisa (CNPq), under process 475389/2008-5,  for partial support to this work.

\bibliography{biblio}

\bibliographystyle{unsrt}

\end{document}